# Bismuth iron garnet $Bi_3Fe_5O_{12}$: a room temperature magnetoelectric material

Elena Popova[*,1], Alexander Shengelaya[2,3], Dimitri Daraselia[2], David Japaridze[2], Salia Cherifi-Hertel[4], Laura Bocher[5], Alexandre Gloter[5], Odile Stéphan[5], Yves Dumont[1], and Niels Keller[1]

[1] Groupe d'Etude de la Matière Condensée (GEMaC), CNRS – UVSQ, Université Paris-Saclay, 78035 Versailles, France

[2] Department of Physics, Tbilisi State University, 0128 Tbilisi, Georgia

[3] Ivane Javakhishvili Tbilisi State University, Andronikashvili Institute of Physics, 0177 Tbilisi, Georgia

[4] Institut de Physique et de Chimie des Matériaux de Strasbourg (IPCMS), CNRS – UNISTRA, 67034 Strasbourg, France

[5] Laboratoire de Physique des Solides (LPS), CNRS – Univ. Paris-Sud, Université Paris-Saclay, 91405 Orsay, France

Abstract

The possibility to control the magnetic properties of a material with an electric field at room temperature is highly desirable for modern applications. Moreover, a coupling between magnetic and electric orders within a single material presenting a wide range of exceptional physical properties, such as bismuth iron garnet (BIG), may lead to great advances in the field of spintronic applications. In particular, the combination of the magnetoelectric (ME) coupling with the low damping of spin waves in BIG can allow the control and manipulation of spin waves by an electric field in magnonic devices. Here we report the unambiguous observation of linear magnetoelectric coupling above 300 K in BIG using ferromagnetic resonance technique with electric field modulation. The measured coupling value is comparable with that observed for prototypal magnetoelectric $Cr_2O_3$. On the basis of our experimental results, the strength of this linear ME coupling is directly linked to the presence of bismuth ions inducing strong spin orbit coupling and to the appearance of local magnetic inhomogeneities related to the magnetic domain structure. The unprecedented combination of magnetic, optical and magnetoelectrical properties in BIG is expected to trigger significant interest for technological applications as well as for theoretical studies.

---

* Corresponding author: olena.popova@uvsq.fr

Materials which allow electric field control of magnetization (or conversely, the control of electric polarization by the application of a magnetic field) via the magnetoelectric (ME) effect have been the subject of investigation for several decades. The recent revival of interest in this field is due to the discovery of new compounds with large ME coupling and novel theoretical concepts [1-3]. The ME effect, when strong enough, could permit the development of novel devices for data storage [4], different kinds of sensors or other applications that require spin-charge conversion. Though the progress in this research field has been significant in the recent years, there are still a number of major challenges to overcome. At present very few compounds with ME coupling at room temperature are known, such as $Cr_2O_3$ [5, 6], $BiFeO_3$ [7, 8] and Z-type hexaferrites [9]. However, $Cr_2O_3$ and $BiFeO_3$ order antiferromagnetically with a very small net magnetization. Z-type hexaferrites are ferrimagnets with a significant magnetization, however, they possess nonlinear ME effects (direct ME coefficient up to a few hundreds $mV.cm^{-1}.Oe^{-1}$ Refs. [9, 10]) and electrical leakage due to residual conductivity. In the absence of suitable single-phase room temperature ME compounds, emphasis has shifted to composite materials where the combination of magnetostrictive and piezoelectric compounds produces electric polarization under the influence of an applied magnetic field [11]. In such composite materials, the largest ME coupling was observed (up to dozens of $V.cm^{-1}.Oe^{-1}$). For garnet-based composite materials the measured direct ME coefficient is typically three orders of magnitude smaller [12]. However, the effect has a resonance character and occurs in a somewhat limited range of frequencies. Therefore, the quest for the single-phase materials with room temperature linear ME effect is still extremely important.

Bismuth iron garnet $Bi_3Fe_5O_{12}$ (BIG) belongs to the Ia-3d cubic space group (Fig. 1a). $Fe^{3+}$ ions occupy tetrahedral (Td) and octahedral (Oh) - with respect to oxygen coordination - sites in the lattice (Fig. 1a, right and left images respectively), and $Bi^{3+}$ ions are located in a dodecahedral sublattice. BIG is ferrimagnetic with a relatively high magnetization of 1600 G at 300 K and magnetic ordering temperature from 650 K to 700 K, depending on Bi content and film thickness [13-15]. The oxygen-stoichiometric BIG is insulating. This material has quite a low spin-wave damping parameter [16]. One of the most useful properties of BIG is the giant Faraday rotation, which makes it suitable for magneto-optical recording and, more recently, for the fabrication of non-reciprocal magneto-optical devices based on magneto-photonic crystals [17, 18]. Furthermore, the existence of energy gaps both at the Fermi level and in the unoccupied spin states [19] is promising for optical and spintronic applications at room temperature, similar to those described in [20].

In this letter we present the evidence of robust room temperature ME coupling in ferrimagnetic bismuth iron garnet thin films using a novel microscopic method for the direct determination of the ME effect based on the standard ferromagnetic resonance (FMR) technique combined with electric field modulation. The linear ME coupling strength was determined quantitatively as a function of temperature for BIG thin films with different cation stoichiometry. This material fulfills most of the requirements anticipated for a single phase ME material in terms of coupling strength above room temperature, large magnetization and low conductivity.

Bismuth iron garnet films were grown on isostructural $Y_3Al_5O_{12}$ (YAG) and $Gd_3Ga_5O_{12}$ (GGG) substrates. Detailed growth and structural information can be found in Refs. [13, 14, 17] and Supplementary Information (SI). Special attention should be given to the control of bismuth transfer from the target to the film, as the film properties are sensitive to the Bi content [15, 17, 21]. To elucidate the influence of cation stoichiometry on film properties, films with different Bi/Fe ratio were studied. The Bi content in the film was estimated from its physical properties [22]. The films are oxygen-stoichiometric and iron valence is solely 3+ (see Fig. 1 in SI) [22]. The cation distribution within the films is uniform and the interface is atomically sharp (Fig. 2 in SI). In the following, experimental results are presented in detail for the cation-stoichiometric BIG/YAG(001) epitaxial system.

Aberration-corrected scanning transmission electron microscopy (STEM) studies confirm the high structural quality of the grown BIG films, as presented in Fig. 1 for a 90 nm thick BIG/YAG(001) film. Figure 1b shows a low-magnified high-angle annular dark-field (HAADF) STEM image of the film at the boundary between two sub-micrometric grains. As for a majority of complex oxides, the growth is columnar with lateral column sizes of around a hundred nanometers. The column boundaries are typically one atomic plane thick and do not contain secondary phases [14], which can be responsible for appearance of, for example, ferroelectricity or other physical phenomena [23]. The fast Fourier transform (FFT) of Fig. 1b (Fig. 1c) confirms the presence of a single phase BIG film grown on the cubic YAG substrate. All films are epitaxial and relaxed for thicknesses above a few tens of nanometers [14]. Atomically resolved HAADF-STEM images in Fig. 1d, e present the BIG structure observed down to the atomic columns. White areas correspond to pure Fe and mixed Bi/Fe atomic columns along the [100] zone axis, as presented with the structure model superimposed in Fig. 1e. White dotted squares in Figs. 1d and 1e correspond to BIG unit cell. The magnetic and magneto-optical properties [22] confirm the high quality and stoichiometry of the BIG/YAG films.

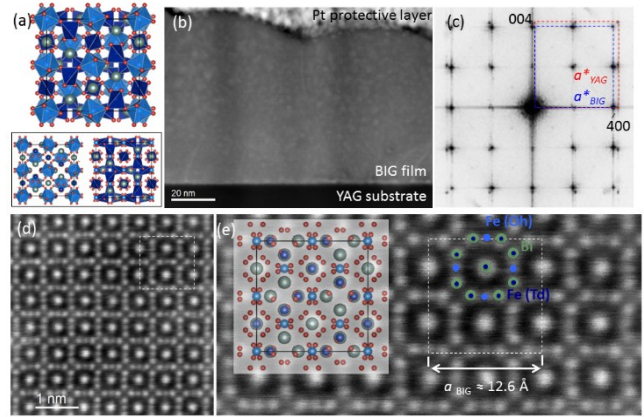

**Figure 1.** BIG atomic structure: (a) Schematic view of BIG unit cell along the [100] direction with the Fe octahedral (Oh) and tetrahedral (Td) sites highlighted in light blue and dark blue, respectively; Bi and O atoms are in green and red respectively. Left (right) inset emphasizes the Oh (Td) sites along the [100] direction. (b) Low-magnified high-resolution HAADF-STEM image of the BIG film grown on YAG(001) substrate. (c) FFT of the image B showing a splitting of the diffraction spots; the BIG and YAG diffraction patterns are highlighted by blue and red dotted lines, respectively. (d) Atomically resolved HAADF-STEM image of the BIG lattice along the [100] zone axis. (e) Enlarged image of the BIG lattice and the corresponding atomic structure. In (d) and (e), the white dotted square represents one BIG unit cell.

ME coupling was measured using the original method combining the standard ferromagnetic resonance technique with electric field modulation (EFM) [24]. The main idea to detect the ME coupling using this method is to apply EFM instead of the usual magnetic field modulation (MFM) to observe FMR signals in ME materials. If the ME coupling is present in the sample, modulation by an electric field, $E(t)$, will lead to a modulation of the magnetization, $M(t)$, and therefore to a modulation of the magnetic field in the sample $B(t)=\mu_0[H+M(t)]$. In this case the FMR signal, which is proportional to the strength of the ME effect, may be detected. The ratio of the FMR signal intensities detected by electric and magnetic modulations is proportional to the ME coupling, which can be determined quantitatively [24]. Details regarding the EFM-FMR experiment can be found in Supplementary Information [22].

Figure 2a shows conventional magnetic field modulated FMR absorption curves recorded at different temperatures with the dc magnetic field perpendicular to the film plane. The inset to this figure represents the variation of the FMR signal intensity, measured as a function of the temperature. This exhibits a classical monotonic temperature dependence of the signal intensity as observed in iron garnets [25]. By switching from magnetic to electric modulation, clear FMR signals were observed, as shown in Fig. 2b. The inset to Fig. 2b shows the temperature dependence of the EFM-FMR signal intensity, exhibiting a pronounced non-monotonic behavior, which reflects the non-monotonic temperature dependence of the ME coupling in BIG, with a maximum around 400 K (see Fig. 2c). It should be noted that the very observation of the electrically modulated FMR signal gives unambiguous proof of the existence of ME coupling in a material, since no signal can be observed in the materials that do not possess ME coupling [22, 24]. Moreover, the ME coupling strength can be determined quantitatively from the ratio of the EFM and MFM-FMR signal intensities. Figure 2c presents the ME coupling strength as a function of temperature. The ME coupling increases from room temperature to about 450 K and then decreases as the temperature approaches magnetic ordering temperature. It was found that the EFM-FMR signal amplitude linearly increases as a function of the applied EFM amplitude, as shown in Fig. 2d. This implies a linear ME effect.

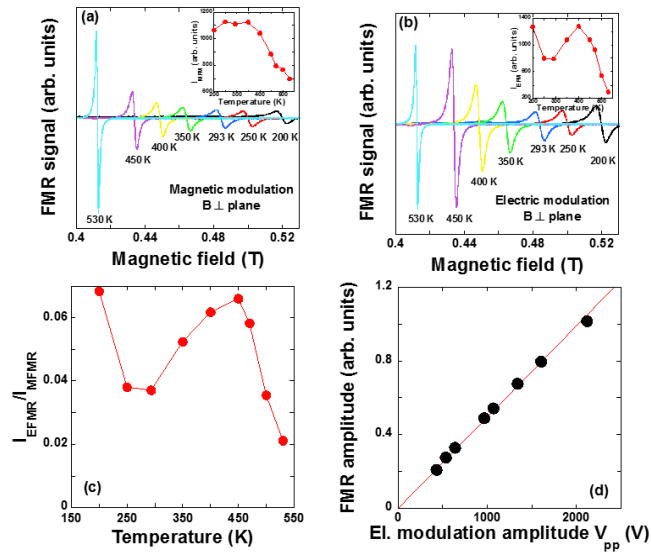

**Figure 2.** Magnetoelectric coupling in BIG/YAG(001) films. (a), (b) Ferromagnetic resonance curves at different temperatures measured using magnetic (electric) field modulation. The insets show the dependence of the integrated signal intensity on temperature. (c) Magnetoelectric coupling strength for BIG/YAG(001) film as a function of the temperature. (d) Variation of FMR line amplitude as a function of the amplitude of electric modulation field.

In these measurements, the so-called converse ME coupling, i.e. the magnetization induced by electric field, can be determined [26]. For the studied film on a YAG(001) substrate (sample A in Table 1), the value of this coupling is determined to be $5\times10^{-6}$ Oe/V/mm at 300 K, which is about 5 times smaller than the maximum value reported for $Cr_2O_3$ at room temperature (around $2\times10^{-5}$ Oe/V/mm [27, 28]). The ME coupling in BIG is observed to be stronger at temperatures below and above room temperature.

Linear magnetoelectric coupling is forbidden in garnets due to symmetry considerations [29-32]. Garnets are thought to preserve their centrosymmetric lattice even under epitaxial strain induced in thin films by substrates. In this case, the stress is accommodated, for example, by changing Fe – O bonds [33, 34]. In fact, garnet films grown on iso-structural substrates with different lattice misfits are generally relaxed above a few tens of nanometers [14]. Linear ME coupling was however observed in quite a few cases of garnet single crystals and epitaxial films [35-40]. A large variety of the measurement techniques have been used to detect the effect in different crystallographic directions within garnets of different compositions. The results are not always easy to compare, however, some trends can be deduced. First of all, the linear ME effect is systematically observed in garnets with a partial cation substitution on different sublattices: $(BiX)_3(FeGa)_5O_{12}$, where X = Lu, Y or YLaPr [37, 38, 40]. The non-uniform distribution of cations affects the cubic symmetry and can, in principle, give rise to the appearance of ME coupling and electric polarization [32]. Recent theoretical work [41] suggests that the lack of space inversion in the dodecahedral magnetic ion sublattice allows ferroelectric ordering in garnets. It should however be stressed that in all the cited experiments, the cation substitution was always accompanied by the addition of Bi in dodecahedral sublattice. The large nuclear charge of Bi leads to the important effect of spin-orbit coupling. More surprising is the presence of linear ME effect in pure and supposedly stoichiometric yttrium iron garnet [35, 36, 39, 42]. Among the explanations, the triclinic YIG symmetry below 130 K [39] or the presence of defects, especially oxygen vacancies [31, 42], are discussed. Indeed, oxygen vacancies lead to the change in iron oxidation state from 3+ to 2+ [31, 34]. The resulting excess electrons – that are engaged in electrical conductivity of a crystal via the hopping mechanism – are thought to freeze below 130 K. These frozen-in dipoles would generate electric field leading to linear magnetoelectric coupling. However, large oxygen off-stoichiometry results in an extremely weak amount of $Fe^{2+}$ [34].

In a more general way, the ME coupling and electric polarization are expected in magnetically inhomogeneous structures [43], such as for example domain walls, or under non-uniform magnetic fields [32]. The micromagnetic structure of a garnet film can be influenced by an electric field, as has been shown by Logginov et al. [37, 38]. The authors have demonstrated the electric field controlled displacement of domain walls in a substituted garnet. As described above, the magnetoelectric effect in garnets can be affected by several phenomena: (i) structural modification with slight deviation from centrosymmetry, due to epitaxial strain or cation substitution on dodecahedral sites, (ii) presence of bismuth ions and strong spin-orbit coupling, (iii) influence of the magnetic domain structure of a sample. In order to clarify different contributions into ME effect, several additional experiments have been performed. The results are summarized in Table 1.

| Sample | Garnet | Cation stoichiometry | MEC at 300 K (Oe/V/mm) |
|---|---|---|---|
| A | BIG | Bi/Fe = 0.6 (Bi = 3) | $5 \times 10^{-6}$ |
| B | BIG | Bi ~ 2.5 | $7 \times 10^{-6}$ |
| C | Nanostructured BIG | Bi ~ 2.3 | $2 \times 10^{-6}$ |
| D | $Bi_2$YIG | Bi ~ 2.1, Y ~ 0.9 | $1 \times 10^{-6}$ |
| E | YIG | Y/Fe ≠ 0.6 | $0.6 \times 10^{-6}$ |

**Table 1.** Sample characteristics: material, cation stoichiometry and value of linear magnetoelectric coupling measured using FMR-EFM.

First of all, one of the possible explanations for the appearance of the ME effect can be the slight lattice deformation due to the epitaxial film growth on a lattice-mismatched substrate. This deformation would lower the cubic symmetry of the garnet and could give rise to magnetoelectrical effects. However, it should be noted that the epitaxial strain, followed by fast lattice relaxation [14], does not lead to the symmetry changes in garnets, therefore this reason for the linear ME effect can be discarded.

Secondly, the influence of the bismuth content on the coupling strength was investigated. Samples B, D and E have different Bi concentrations, as shown in Table 1. The coupling strength decreases with decreasing Bi content. The decrease of ME coupling, with respect to pure BIG films (samples A and B), is observed in BIG films partially substituted by yttrium (sample D) and in non-stoichiometric single-crystalline yttrium iron garnet film (sample E). This leads us to the conclusion that the ME coupling enhancement is due to the increased bismuth content of the film rather than the symmetry changes due to the partial cationic substitutions on the dodecahedral sublattice.

Finally, linear ME coupling has also been observed in a nanostructured BIG sample C (for the details concerning this sample see [22]). The magnetoelectric coupling value is lower than that of the continuous sample B (Table 1). This result can be explained by differences between the contour and density of domains in continuous films as compared to patterned samples.

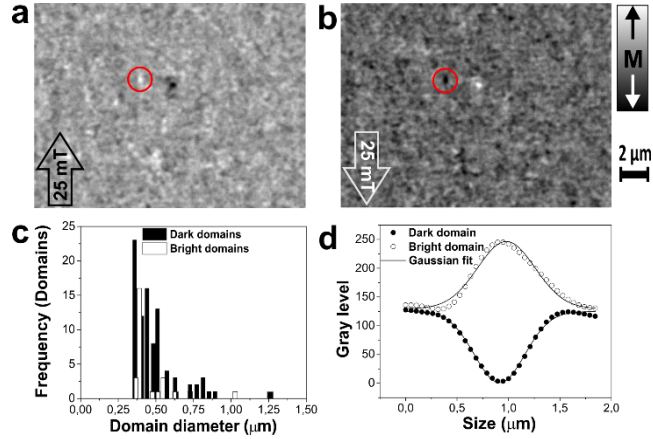

**Figure 3.** Continuous film BIG/GGG(001). Magnetic domain configuration in positive (a) and negative (b) applied field. The corresponding domain analysis showing (c) the frequency and size of the domains with magnetization M collinear to the applied magnetic field. Panel (d) displays line profiles of selected domains, highlighted by a red circle in images (a) and (b). The profiles are fitted with a Gaussian function (continuous line) with an averaged full width at half maximum of 0.5 μm.

Figures 3a-b show the magnetic domain structure in a continuous 300 nm thick film of BIG/GGG(001) observed by means of magneto-optical Kerr microscopy with oppositely applied magnetic field directions. A statistical analysis of the domain size and shape has been performed based on the digital images by applying suitable threshold values of the grayscale level (Fig. 3c). The domains have a predominantly circular shape and a typical diameter of less than 0.5 μm (Fig. 3d). This could be ascribed to the nanometric crystallographic domain size, as the oxide growth is columnar [14]. In contrast, the domain structure in patterned films depends on the shape and size of the microstructures. In the case of micro-stripes designed in the studied continuous BIG/GGG(001) film, micrometric domains develop, with alternated domain structures or magnetic flux closure configurations [22]. The domain wall density in the continuous film is about twice that of the patterned film. The reduced quantity of domain walls in the patterned sample is consistent with a lower ME coupling value as compared with continuous films.

In summary, unexpected first order magnetoelectric effect – incompatible with centrosymmetric lattice – was observed in cation-stoichiometric bismuth iron garnet. This ME effect is detected for temperature well above 300 K. The ME coupling strength is comparable to that of the prototype antiferromagnetic $Cr_2O_3$. Several possible explanations of this phenomenon have been considered in the present work. It has been demonstrated that this ME coupling is not associated with strain induced by the lattice misfit with the substrates. The ME coupling strength in BIG depends on two main contributions. First of all, the magnetoelectric coupling is strongly related to the presence of local magnetic inhomogeneities, such as, for example, domain walls or vortexes. Domain wall density plays undoubtfully an important role in the strength of the coupling. Secondly, we have demonstrated, using EFM-FMR, that the magnetoelectric coupling in the cation off-stoichiometric garnets is significant, increasing notably in Bi-containing non-mixed garnet. Therefore, the role of bismuth ions, that have strong spin-orbit coupling, is as important as the domain structure of the material and should be considered in future theoretical work on the subject. The enhancement of the magnetoelectric effect in bismuth iron garnet with respect to yttrium iron garnet and the non-monotonic dependence of the ME coupling on the temperature raise additional questions about the mechanisms of magnetoelectric interaction in garnets. This enhanced ME effect is of great importance for the rapidly developing field of magnonic devices, where it can be used for highly desirable electrical tuning of spin waves [44, 45].

**Supporting Information**

Additional supporting information may be found in the online version of this article at the publisher's website.

**Acknowledgements**

The authors acknowledge financial support from the CNRS-CEA "METSA" French microscopy network for the STEM-EELS experiments and the Ile de France "C'Nano - IdF" network for magnetic measurements (NOVATECS program n°IF-08-1453/R). This work was also supported by the Georgian National Science Foundation Grant No. RNSF/DI/21/6-160/14. The sample patterning has been performed in IEF (Orsay) by L. Magdenko and B. Dagens and in CRISMAT (Caen) by M. Strebel and X. Larose.